\documentclass[12]{article}

\usepackage{amsmath,amssymb,amsfonts}
\usepackage{mathrsfs}
\usepackage{graphicx}
\usepackage{subfig}
\usepackage{float}
\usepackage{physics}
\usepackage{authblk}
\usepackage[left=2cm,right=2cm,top=2cm,bottom=2cm]{geometry}
\usepackage{hyperref}
\hypersetup{colorlinks = true,
	linkcolor = red,
	anchorcolor = blue,
	citecolor = blue,
	filecolor = blue,
	urlcolor = blue}

\DeclareMathOperator{\arctanh}{arctanh}

\begin{document}
\author[1]{Braulio M. Villegas-Martínez}
\author[1]{Francisco Soto-Eguibar\thanks{Corresponding author: feguibar@inaoep.mx}}
\author[2]{Sergio A. Hojman}
\author[3]{Felipe A. Asenjo}
\author[1]{Héctor M. Moya-Cessa}
\affil[1]{Instituto Nacional de Astrofísica Óptica y Electrónica.
Calle Luis Enrique Erro No. 1, Santa María Tonantzintla, Pue., 72840, Mexico}
\affil[2]{Departamento de Ciencias, Facultad de Artes Liberales,
Universidad Adolfo Ibáñez, Santiago, 7491169, Chile

Departamento de Física, Facultad de Ciencias, Universidad de
Chile, Santiago, 7800003, Chile

Centro de Recursos Educativos Avanzados, CREA, Santiago,
7500018, Chile.}
\affil[3]{3Facultad de Ingeniería y Ciencias, Universidad Adolfo Ibáñez, Santiago, 7491169, Chile.}

\title{Non-unitary transformation approach to $\mathcal{PT}$ dynamics}

\maketitle

\begin{abstract}
We show that several Hamiltonians that are $\mathcal{PT}$ symmetric may be taken to Hermitian Hamiltonians via a non-unitary transformation and vice versa. We  also show that for some specific Hamiltonians such non-unitary transformations may be associated, via a fractional-Wick rotation, to complex time.
\end{abstract}

\section{Introduction}\label{sec1}
Non-Hermitian Hamiltonian systems have become a multifaceted research frontier which encompasses a wide range of theoretical as well as experimental interdisciplinary fields \cite{1,1.1,2,3,4,5,6}. The choice of the proper framework to investigate the non-Hermitian Hamiltonians depends on the nature of their eigenvalues. A particular emphasis of interest centers on a class of these systems obeying the combined parity-time ($\mathcal {PT}$) symmetry. The main reason is that this generalization of the conventional quantum mechanics to the complex domain has opened an alternative condition (but not sufficient) to guarantee the reality of the spectrum for non-Hermitian Hamiltonians \cite{7,8,9,fring}. This field of research traces back to Bender's original conjecture \cite{7,8,9,bender98,bender02,12}  where the hermiticity condition can be relaxed under the assumption of space-time reflection symmetry; soon after, a more general theoretical proposal than the $\mathcal{PT}$ concept was given in \cite{12.1,12.2,12.3}. Naturally, Bender's pioneering work has marked the grounds for these non-Hermitian Hamiltonians to acquire a new meaning with $\mathcal {PT}$-symmetry in non-Hermitian quantum mechanics \cite{13,14,15,16}. These concepts indeed have already led to compelling applications in compound photonic structures, for example $\mathcal {PT}$-symmetric Hamiltonians found an ideal paragon on coupled waveguides differentiated by absorption and amplification into their refractive index profiles \cite{17,18,19,20,21,21.1,22}. On the other hand, in the light of elementary quantum theory, the acceptance of $\mathcal {PT}$-symmetric Hamiltonians have been flourishing for a new framework where can be mapped into a Hermitian ones by under specific transformations that underlie the reality of their spectrum. So far most effort has gone the grounds for treating time-independent and the time-dependent scenarios of such classes of Hamiltonians through Dyson maps, gauge-like transformations, Bogoliubov, Darboux or non-unitary transformations \cite{23,24,25,25.1,25.2,25.3,25.4}; most of them, however, involve cumbersome procedures and complicated requirements, which currently limits the scope of their practical applications. Following the seminal insight of these developments, we hence proceed in this work to look for non-unitary transformations that applied to Hermitian Hamiltonians, i.e. Hamiltonians that have real eigenvalues, produce $\mathcal{PT}$-symmetric Hamiltonians, and vice versa. Broadly speaking, our purpose is not to engage in a comprehensive review; rather, we want to show that under an adequate choice of the arguments of the non-unitary transformations, it is possible to produce some renown $\mathcal{PT}$-symmetric Hamiltonians. Resulting non-unitary transformations offer new perspectives, and are substantially simpler than those reported above in literature. Particularly, we show that one of those transformations may be associated with a fractional-Wick rotation to complex time.

\section{Non-conservative binary oscillator}\label{sec2}
\begin{figure}[H]
\centering
\subfloat[Complex plot  of $\sqrt{k^2-\gamma^2}$]
{\includegraphics[width=0.45 \textwidth]{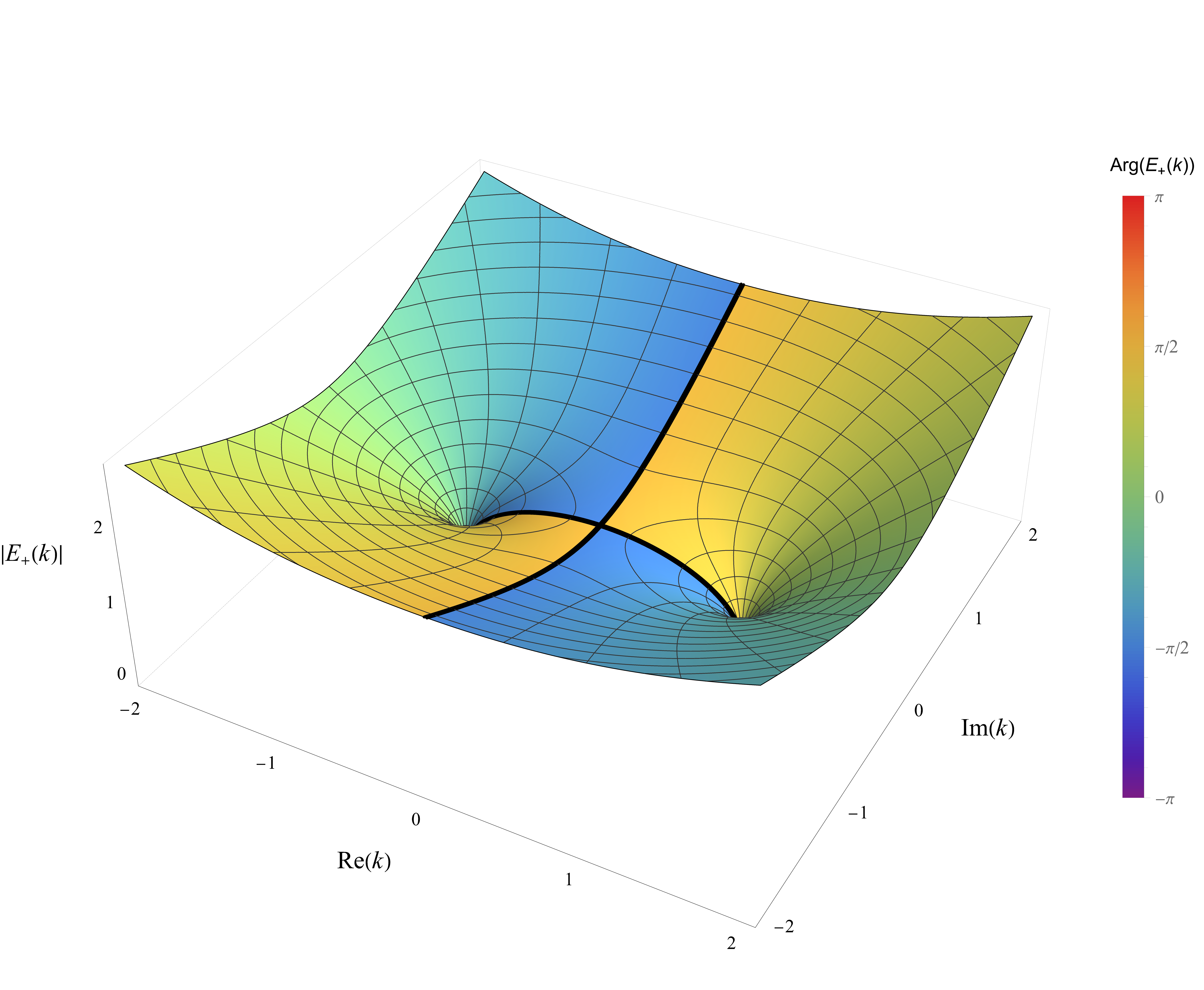}}
\qquad
\subfloat[$\arg\left( \sqrt{k^2-\gamma^2}\right) $]
{\includegraphics[width=0.45\textwidth]{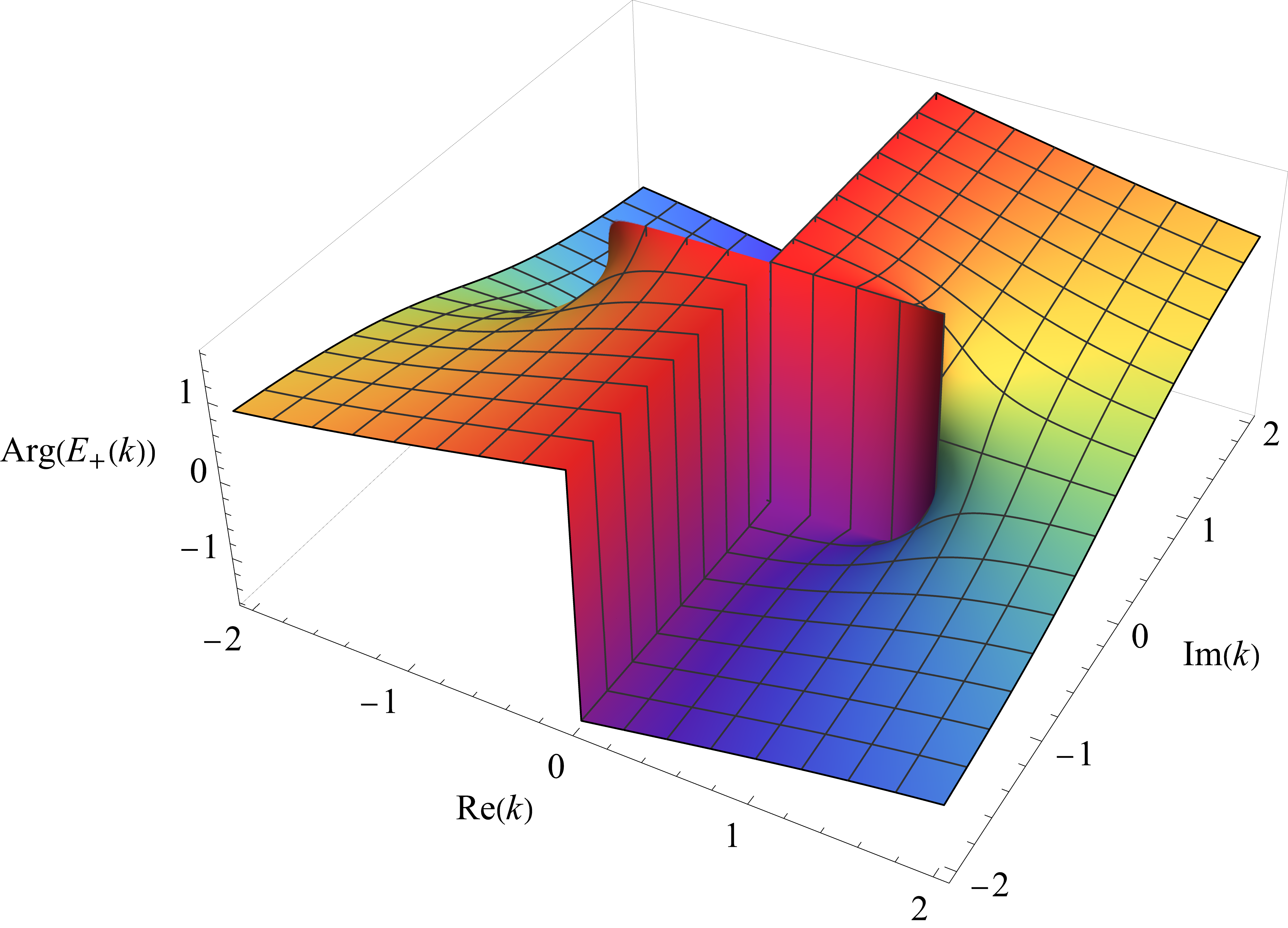}} 
\caption{The positive branch of $\sqrt{k^2-\gamma^2}$ as function of $k$ complex for $\gamma=1$.}\label{f0020}
\end{figure}

\begin{figure}[H]
\centering
\subfloat[Complex plot  of $-\sqrt{k^2-\gamma^2}$]
{\includegraphics[width=0.45 \textwidth]{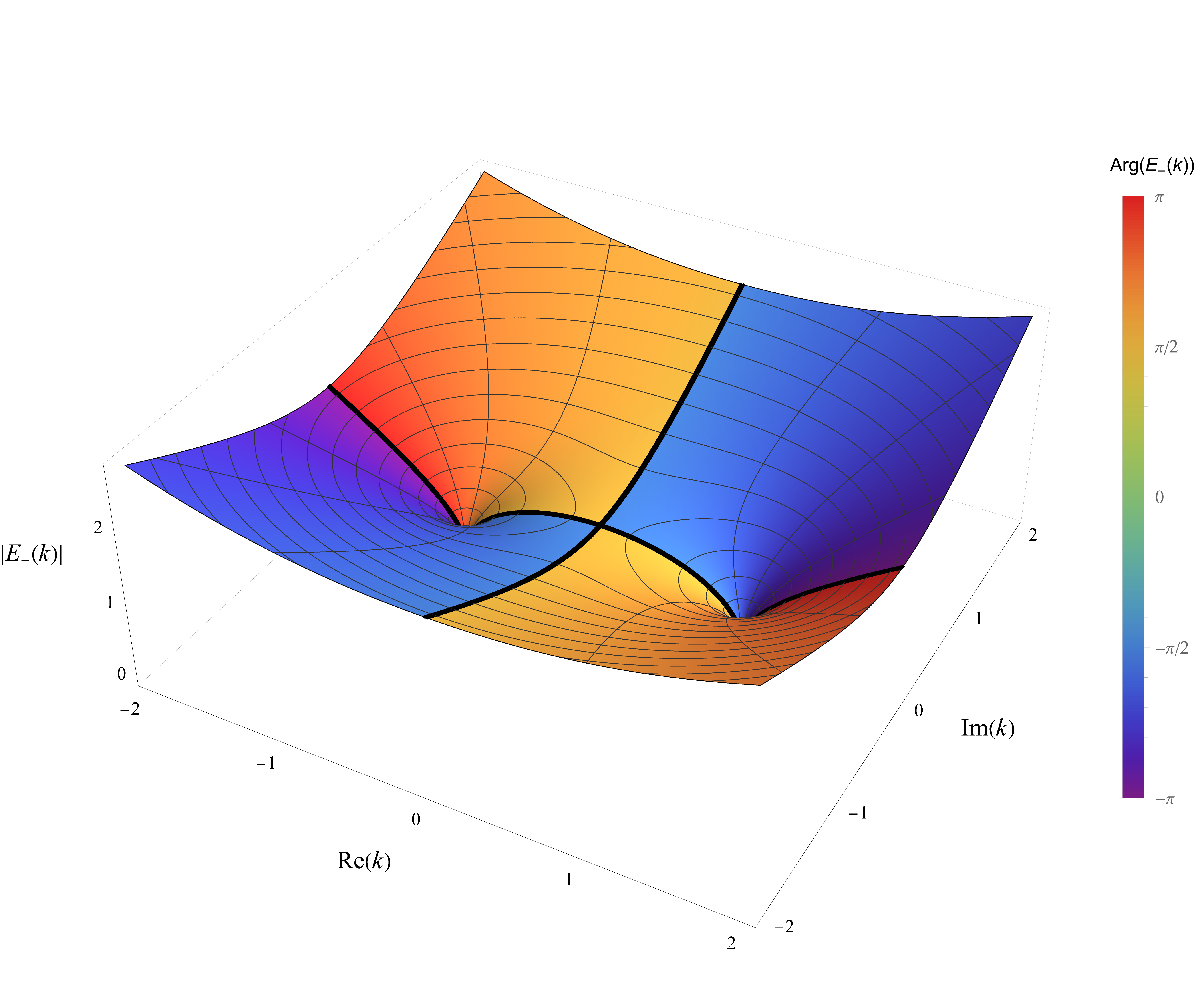}}
\qquad
\subfloat[$\arg\left( -\sqrt{k^2-\gamma^2}\right) $]
{\includegraphics[width=0.45\textwidth]{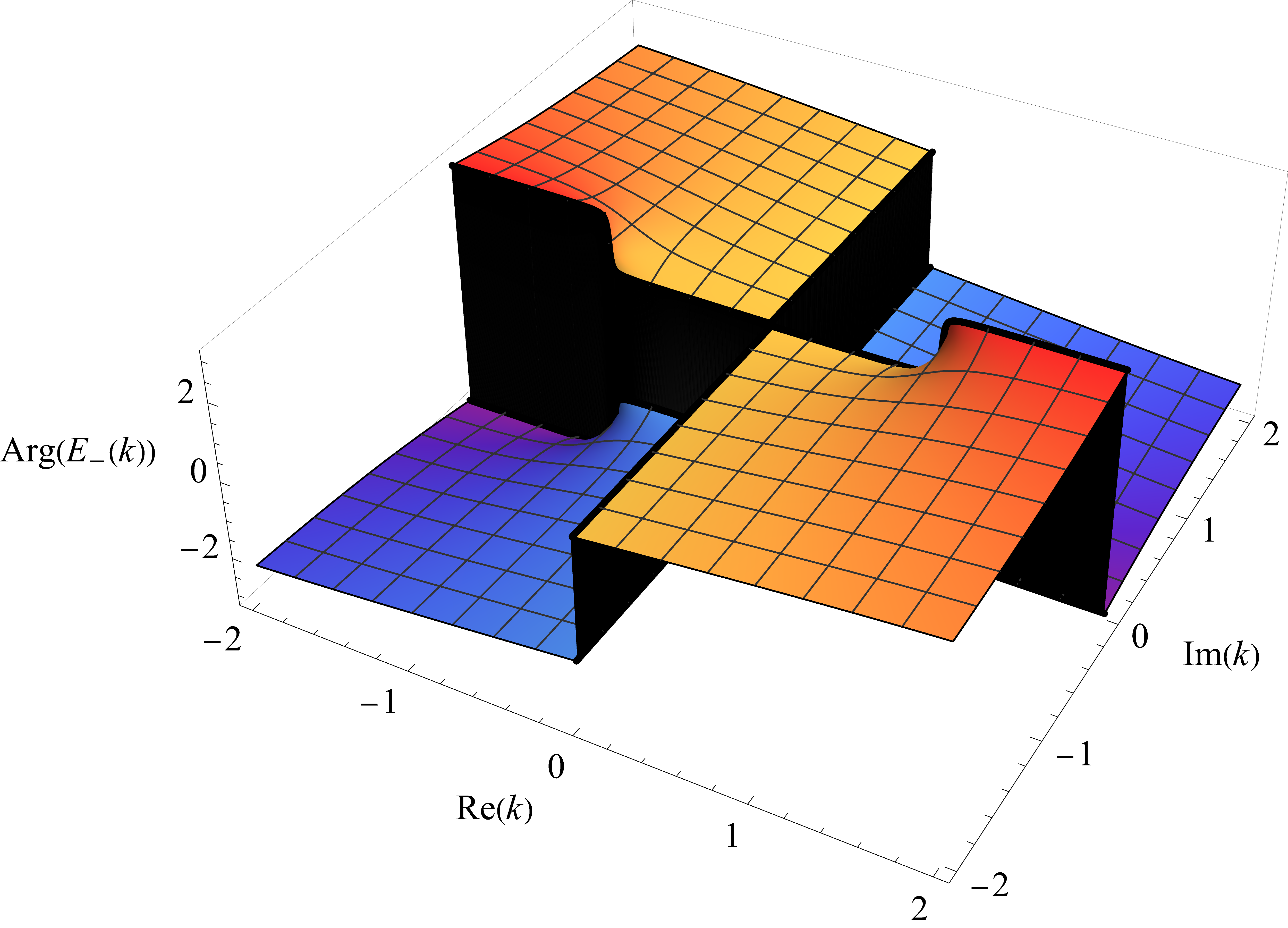}} 
\caption{The negative branch of $\sqrt{k^2-\gamma^2}$ as function of $k$ complex for $\gamma=1$.}\label{f0030}
\end{figure}

We consider first the $\mathcal{PT}$ symmetric Hamiltonian used in \cite{NatComm},
\begin{equation}\label{0010}
\hat{H}^{(\mathcal {PT})}=\epsilon 1_{2\times 2}+k\sigma_x+i\gamma\sigma_z,
\end{equation}
that describes two equally tuned oscillators at energy level $\epsilon$ and with their attenuation rates differing by $2\gamma$. In this Hamiltonian $\sigma_x,\,\sigma_y,\,\sigma_z$ are the Pauli spin matrices, $1_{2\times 2}$ is the unit $2\times 2$ matrix, and $\epsilon, \, k, \, \gamma$ are arbitrary real constants; we will also assume, for simplicity and without loss of generality, that $k$ and $\gamma$ are strictly positive. Notice that the Hamiltonian \eqref{0010} is $\mathcal {PT}$ symmetric but it is not Hermitian.\\
The Hamiltonian \eqref{0010} may be transformed via $\exp(-\theta \sigma_y)$ to
\begin{equation}\label{0020}
\hat{H}^{(\mathcal {PT})} = \exp(-\theta\sigma_y)  \hat{H}   \exp(\theta\sigma_y)
\end{equation}
with
\begin{equation}\label{0030}
\hat{H}=\epsilon 1_{2\times 2}+\alpha\sigma_x.
\end{equation}    
We may obtain the angle of {\it rotation} from
\begin{align}\label{0040}
\exp(-\theta\sigma_y)&\hat{H}\exp(\theta\sigma_y)=\epsilon 1_{2\times 2}
+\alpha[\sigma_x \cosh (2\theta)+i\sigma_z\sinh(2\theta)],
\end{align}
which establishes the conditions $\alpha\cosh (2\theta)=k$ and $\alpha\sinh (2\theta)=\gamma$; among the infinite number of solutions of these equations, for simplicity we choose for $k\neq\gamma$,
\begin{equation}\label{0050}
\theta=\frac{1}{2} \ln \left( \frac{k+\gamma}{\sqrt{k^2-\gamma^2}} \right),  \qquad \alpha=\sqrt{k^2-\gamma^2}, \qquad | k | \neq  \gamma 
\end{equation}
for $k=\gamma$, we get $\tanh(2\theta)=1$ that does not have a solution and the transformation $\exp(-\theta\sigma_y)$ breaks down. As we will see below, this corresponds to the exceptional points of the $\mathcal {PT}$ symmetric Hamiltonian.\\
The two eigenvalues of $\hat{H}$, \eqref{0030}, are $\epsilon\pm \alpha$. These two eigenvalues are real as long as $\alpha$ is real; which, of course, also establishes the region for which the Hamiltonian is Hermitian. For the $\mathcal{PT}$ symmetric Hamiltonian $ \hat{H}^{(\mathcal {PT})} $, \eqref{0010}, the eigenvalues are $\epsilon \pm \sqrt{k^2-\gamma^2}$.\\
We then have three regions:
\begin{enumerate}
\item When $k>\gamma$, the difference between the attenuation rates of the two equally tuned oscillators must be smaller than the natural frequency of the oscillators. In this case, $\theta $ and $\alpha$ are real, the Hamiltonian $\hat{H}$, \eqref{0030}, is Hermitian, and so its two eigenvalues are real. The transformation $ \exp(-\theta\sigma_y) $ is not unitary. The two eigenvalues of the $\mathcal{PT}$ symmetric Hamiltonian $ \hat{H}^{(\mathcal {PT})} $, \eqref{0010}, are also real, and we are in the unbroken symmetry region.
\item When $k<\gamma$, the \textit{angle} $\theta$ becomes complex and the transformation $ \exp(-\theta\sigma_y) $ consists of the composition of a unitary transformation and a non unitary one. The parameter $\alpha$ also becomes complex, and the eigenvalues of both Hamiltonians, $ \hat{H}^{(\mathcal {PT})} $ and $ \hat{H} $ are complex numbers. In this case, we are in the broken symmetry region.
\item When $k=\gamma$, the corresponding matrices to $ \hat{H}^{(\mathcal {PT})} $ and $ \hat{H} $ are defective; we are, in this case, in the exceptional points. At this value of the parameters, the eigenvalues coalesce and there is crossing between the energy levels. At these points there is only one eigenvalue, $\epsilon$, and then the corresponding Hamiltonian matrix becomes defective, which means that it is non diagonalizable.
\end{enumerate}
The behavior of the eigenvalues as function of the real parameter $k$, described in the previous points is depicted in Fig~\ref{0010}.
\begin{figure}[H]
\centering
\includegraphics[width=0.9\linewidth]{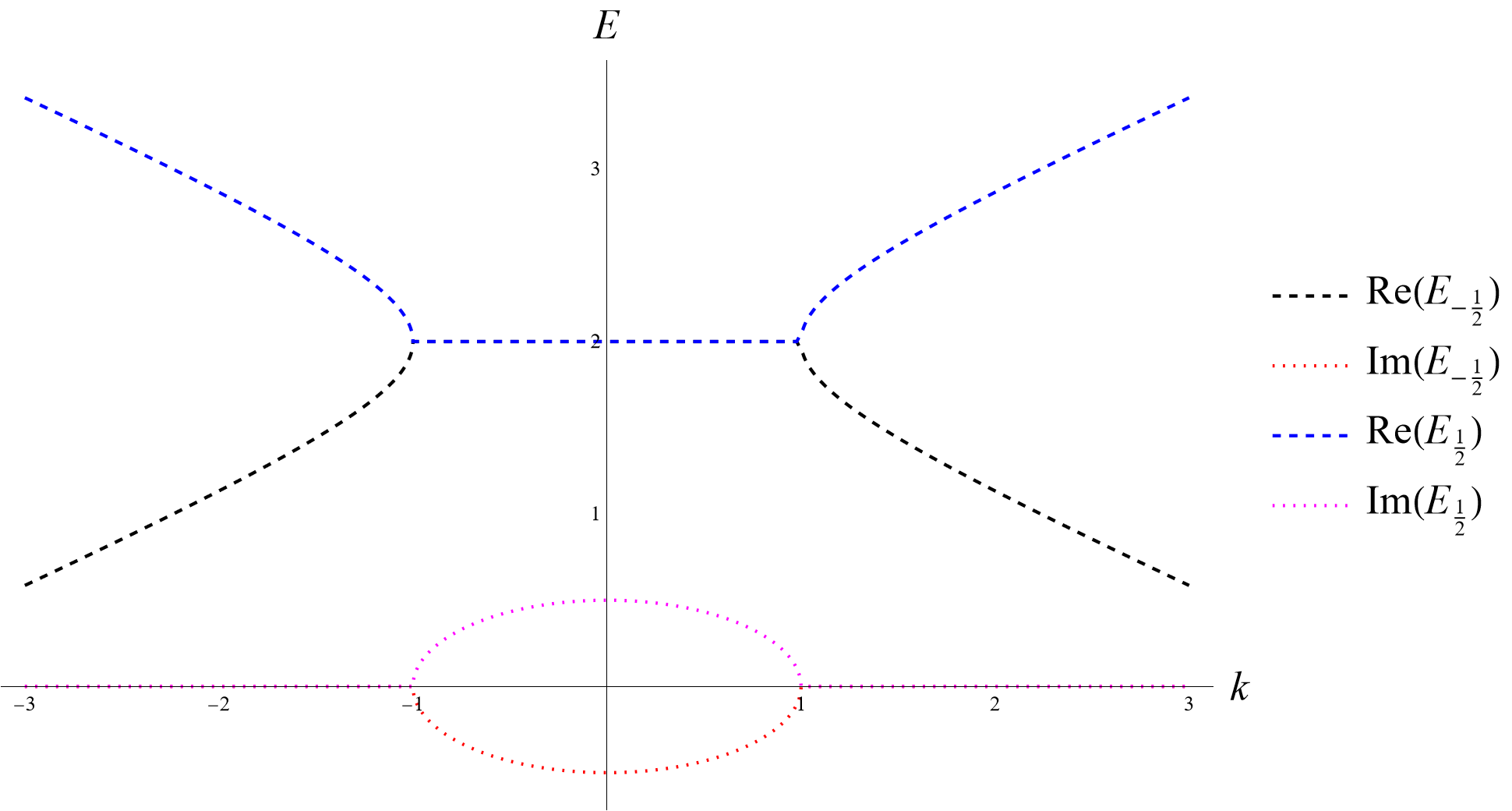}
\caption{The real and imaginary parts of the energy eigenvalues as function of the parameter $k$ for $\epsilon=2$ and $\gamma=1$.}
\label{f0010}
\end{figure}
From the mathematical point of view, the relative complexity of the behavior of the eigenvalues comes from the fact that the function $\sqrt{k^2-\gamma^2}$ is multi-valued and therefore has branch points; as explained by Bender \cite{16}, we get a complete picture of the behavior of the eigenvalues when we make the complex deformation of the problem. As $\epsilon$ is just an additive constant, it will be enough to analyze the behavior of the multivalued function $\sqrt{k^2-\gamma^2}$, when we extend the interaction parameter $k$ to become complex.\\
In Figure \ref{f0020}, we present the \textquotedblleft positive\textquotedblright \, branch of $\sqrt{k^2-\gamma^2}$ as function of $k$, extending this variable to the complex domain. We observe the branch points, that in the left subfigure are marked in black; in the right subfigure we can view the discontinuity in the argument of the function.
In Figure \ref{f0030}, we present the \textquotedblleft negative\textquotedblright \, branch of $\sqrt{k^2-\gamma^2}$ as function of $k$, as a complex variable. Exactly as in the other branch, we have marked the branch points in black in the left subfigure; also the discontinuity of the argument of the function can be seen in the right subfigure.

\section{$\hat{J}_x$ lattice}\label{sec3}
\begin{figure}[H]
\centering
\includegraphics[width=0.9\linewidth]{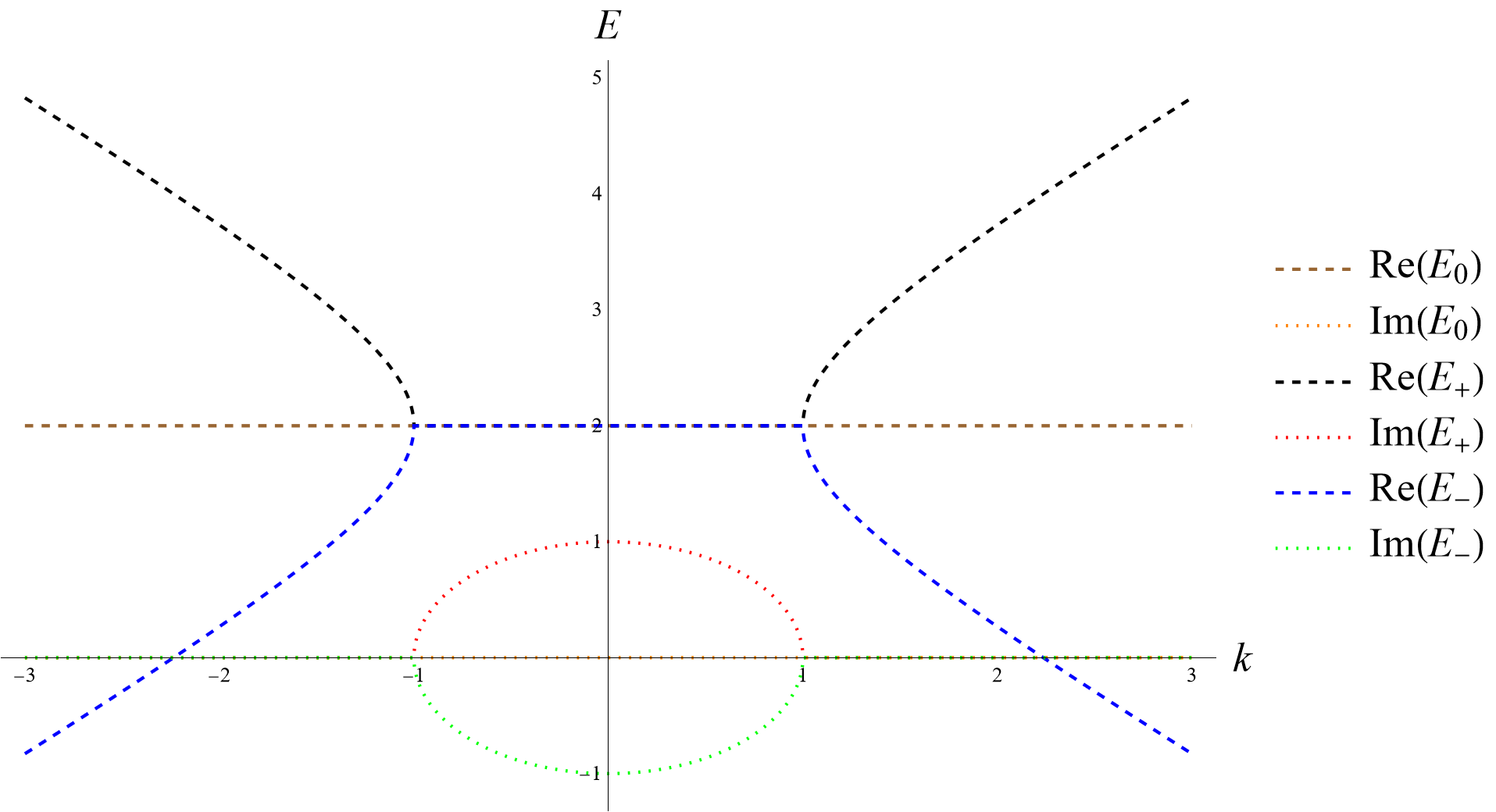}
\caption{The real and imaginary parts of the energy eigenvalues  for $j=1$ as function of the parameter $k$ for $\epsilon=2$ and $\gamma=1$.}
\label{f0040}
\end{figure}

\begin{figure}[H]
\centering
\includegraphics[width=0.9\linewidth]{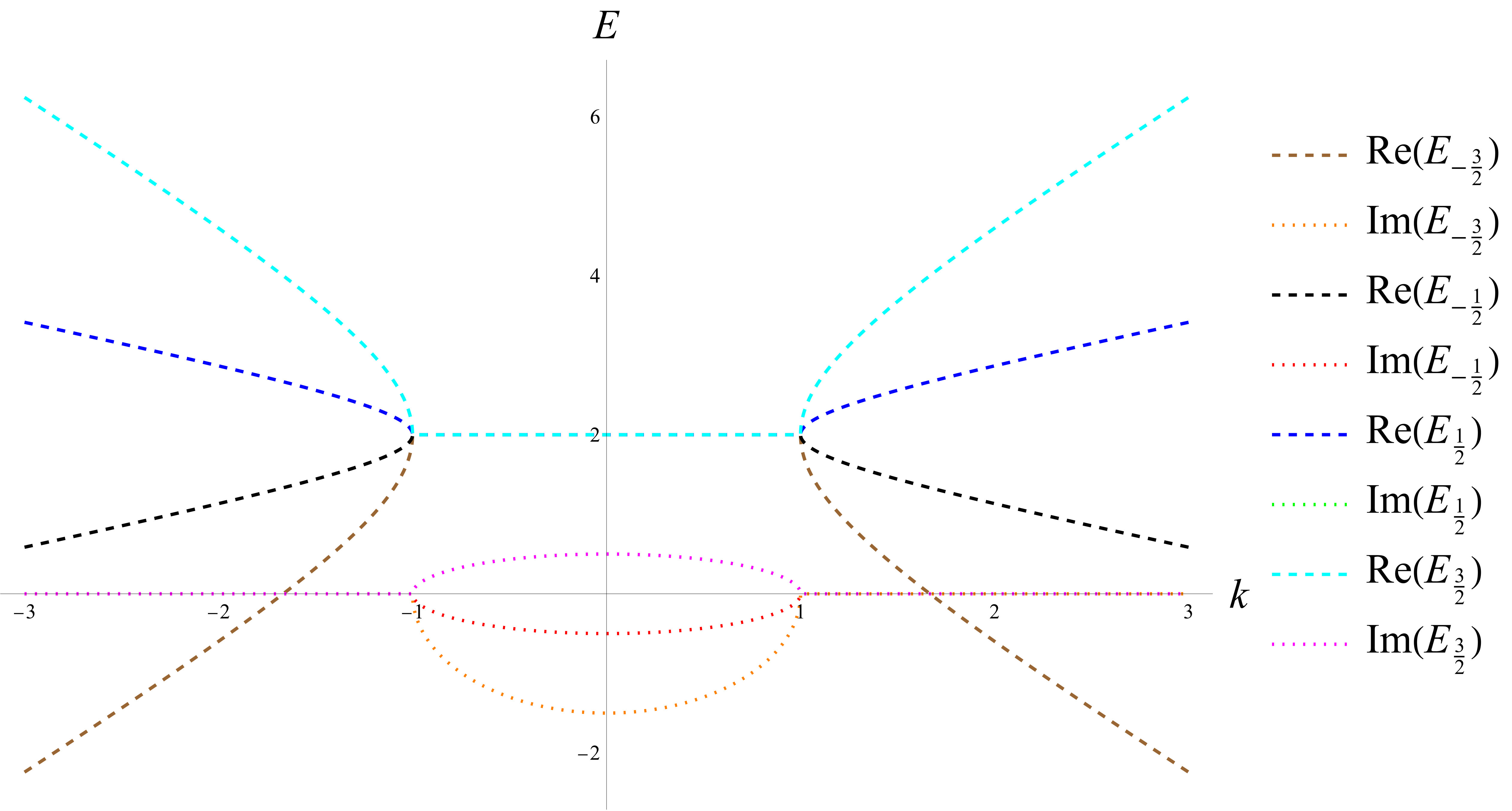}
\caption{The real and imaginary parts of the energy eigenvalues for $j=3/2$ as function of the parameter $k$ for $\epsilon=2$ and $\gamma=1$.}
\label{f0050}
\end{figure}

Let us consider the angular momentum operators $\left\lbrace \hat{J}_x,\hat{J}_y,\hat{J}_z\right\rbrace $, defined by the usual commutations rules \cite{zettili,griffiths,hallmc}
\begin{equation}\label{0120}
\left[\hat{J}_x,\hat{J}_y \right] =i \hat{J}_z, \qquad
\left[\hat{J}_y,\hat{J}_z \right] =i \hat{J}_x, \qquad
\left[\hat{J}_z,\hat{J}_x \right] =i \hat{J}_y.
\end{equation}
We introduce the Hamiltonian \cite{JX}
\begin{equation}\label{0130}
\hat{H}^{\left( \mathcal{PT} \right)} =\epsilon \hat{I}+ i \gamma \hat{J}_z+ k \hat{J}_x,
\end{equation}
with $\epsilon,\;\gamma$ and $k$ reals.\\
Note that:
\begin{enumerate}
  \item The Hamiltonian $\hat{H}^{\left( \mathcal{PT} \right)}$, Eq.~\eqref{0130}, is not Hermitian (symmetric).
  \item The Hamiltonian $\hat{H}^{\left( \mathcal{PT} \right)}$, Eq.~\eqref{0130}, is $\mathcal{PT} $ symmetric when the parity transformation $\mathcal{P}$  is the conjugated transposition.
  \item The first part of the Hamiltonian $\hat{H}^{\left( \mathcal{PT} \right)}$, $\epsilon \hat{I}+ i \gamma \hat{J}_z$ , is diagonal and can be considered as an ensemble of systems without interaction or as one system with several energetic levels that don't interact either. The second part of the Hamiltonian, $k \hat{J}_x$, introduces an interaction between the elements that compose it or between the levels that it has. Thus, we can consider $k$ as the parameter of interaction \cite{16}.
\end{enumerate}
We make now the transformation
\begin{equation}\label{0140}
    \hat{H}=\exp\left(\theta \hat{J}_y\right) \hat{H}^{(\mathcal{PT})} \exp\left(-\theta \hat{J}_y\right),
\end{equation}
where the parameter $\theta$ can be a complex number, and applying the Hadamard's lemma \cite{hall,miller,rossman}, we obtain
\begin{equation}\label{0150}
    \hat{H}=\epsilon \hat{I}+i\left[\gamma \cosh\left(\theta \right)-k \sinh\left(\theta \right)  \right]\hat{J}_z +\left[ k \cosh\left(\theta \right)- \gamma \sinh\left(\theta \right)\right] \hat{J}_x.
\end{equation}
We choose the parameter $\theta$ in such a way that the coefficient of $\hat{J}_z$ above be zero; thus,
\begin{equation}\label{0160}
\theta_0=\ln \left(\sqrt{\frac{k+\gamma}{k-\gamma}} \right),
\end{equation}
and
\begin{equation}\label{0170}
    \hat{H}_0=\epsilon \hat{I}+\sqrt{k^2-\gamma^2} \hat{J}_x.
\end{equation}
Note that there are an infinite number of ways to choose the $\theta$ parameter so as to make the coefficient of $\hat{J}_z$ zero. We have arbitrarily chosen the one given by Eq.~\eqref{0160} because it is the one that best suits us for simplicity; however, we could have chosen any other of the possible ones and the essence of the result would be exactly the same.\\
Remark also that:
\begin{enumerate}
\item The Hamiltonian \eqref{0170} is Hermitian if and only if  $ | k | \geq  \vert \gamma \vert  $; note that in that case, $\theta_0$ will be real. 
\item If $|k|<\vert\gamma\vert$, $ \sqrt{k^2-\gamma^2}$ will be a complex number and the Hamiltonian \eqref{0170} is not Hermitian; however, it is $\mathcal{PT} $ symmetric when the parity transformation $\mathcal{P}$  is defined as the conjugated transposition.
\item When $|k|=\vert\gamma\vert$, the Hamiltonian \eqref{0170} becomes trivial and the $\theta$ parameter is undefined. As we will see below, this case corresponds to the exceptional points.
\end{enumerate}
The spectrum of the Hamiltonian \eqref{0130} is
\begin{equation}\label{0180}
E_l=\epsilon+l \sqrt{k^2-\gamma^2}, \quad l=-j,...,+j;  \qquad j=\frac{1}{2},1,\frac{3}{2},2,\dots, \quad N=2j+1,
\end{equation}
where $N=2j+1$ is the dimension of the space. Note that the case $j=\frac{1}{2}$ reduces to the non-conservative binary oscillator of the previous section.\\
As expected, if $|k|\geq\vert\gamma\vert$ the spectrum is real as such as the Hamiltonian is Hermitian; however, when $|k|=\vert\gamma\vert$ the parameter $\theta$ is undefined, the spectrum consists of only one value, $\epsilon$, and the Hamiltonian becomes defective, i.e., non-diagonalizable. As we will see below, this case corresponds to the exceptional points.\\
If we remove the restriction that $|k|\geq\vert\gamma\vert$, things get more interesting because we will have three regions. First, when $|k|>\vert\gamma\vert$ we will have the unbroken symmetry region; second, when $|k|<\vert\gamma\vert$ we will have the broken symmetry region; and finally and third, when  $|k|=\vert\gamma\vert$ we will have the exceptional points.\\
In the unbroken symmetry region, i.e., when $|k|>\vert\gamma\vert$, the spectrum is real, the eigenstates oscillates and the systems, or levels of the system, remains in equilibrium between them.\\
In the broken symmetry region, i.e., when $ | k | < \vert\gamma\vert$, the spectrum is complex, eigenvalues appear in pairs, being complex conjugates of each other. Some of the eigenstates grow and others decay; the systems, or levels of the system, are not in equilibrium.\\
As we already mentioned, when $|k|=\vert\gamma\vert$, we get the exceptional points. At this value of the parameters, the eigenvalues coalesce and there is crossing between the energy levels. At these points there is only one eigenvalue, $\epsilon$, and then the corresponding Hamiltonian matrix becomes defective, which means that it is non diagonalizable.\\
Exactly in the same way that the case of the previous section, the complicated behavior comes from the fact that the function $\sqrt{k^2-\gamma^2}$ is multi-valued and therefore has branch points. The arguments presented in Figure \ref{f0010} are the same for all dimensions.\\
In Figure \ref{f0040}, we plot the real and imaginary parts of the energy eigenvalues for $j=1$ (dimension 3) as function of the parameter $k$ for $\epsilon=2$ and $\gamma=1$. In this case we have three eigenvalues, $\left\lbrace  \epsilon-\sqrt{k^2-\gamma^2},\epsilon,\epsilon+\sqrt{k^2-\gamma^2}\right\rbrace$. In this case again the non-broken symmetry region is the region where $|k|>1$, the broken symmetry region is the one where $|k|<1$ and the exceptional points are $k=-1$ and $k=1$.It is worth mentioning that we have three eigenvalues coalesce at $k=1,-1$ and the system exhibits a third-order exceptional point. In the unbroken symmetry region, all the eigenvalues are real. In the broken symmetry region, two of the eigenvalues are complex, and one is the complex conjugate of the other; the other eigenvalue, $\epsilon$ is always real.
In Figure \ref{f0050}, we plot the real and imaginary parts of the energy eigenvalues for $j=3/2$ (dimension 4) as function of the parameter $k$ for $\epsilon=2$ and $\gamma=1$. In this case we have four eigenvalues,
$$ \left\lbrace \epsilon-\frac{3}{2}\sqrt{k^2-\gamma^2},\epsilon-\frac{1}{2}\sqrt{k^2-\gamma^2},\epsilon+\frac{1}{2}\sqrt{k^2-\gamma^2},\epsilon+\frac{3}{2}\sqrt{k^2-\gamma^2}\right\rbrace.$$
We observe that the eigenenergies are real for $|k|>1$ as corresponds to the unbroken symmetry region; for $|k|<1$, we are in the broken symmetry region and the eigenergies are complex and become in pairs; at the exceptional points, again $k=-1$ and $k=1$, the eigenenergies merge. 
In this case, four eigenvalues coalesce sharing a single eigenstate, this is a simple example of a system which exhibits a four-order exceptional point; such behavior it can be observed from the energy bifurcation diagram of Fig.\ref{f0050}. Therefore, the nature of the dimension $N=2j+1$ of the Hamiltonian \eqref{0130} leads to the existence of higher order exceptional points.

\section{The complex deformed harmonic oscillator}\label{sec4}
A very well known $\mathcal{PT}$ symmetric Hamiltonian, but not Hermitian, is the complex deformed harmonic oscillator Hamiltonian given by
\begin{equation}
\hat{H}= \hat{p}^2+\hat{x}^2(i\hat{x})^{\epsilon},
\end{equation}
where $\epsilon$ is a non-negative real number \cite{bender98,bender02,16}.\\
We make the transformation $\ket{\psi}=\hat{S}_r^{-1}\ket{\psi}$, with $\hat{S}_r=\exp\left[\frac{r}{2}\left(\hat{x}\hat{p}+\hat{p}\hat{x}\right)\right]$, such that the transformed Hamiltonian is
\begin{equation}
   \hat{{\mathcal H}}=\hat{S}_r\hat{H}\hat{S}_r^{-1}.
\end{equation}
We can write explicitly the previous Hamiltonian using the expressions
\begin{subequations}
\begin{align}
\hat{S}_r\hat{x}\hat{S}_r^{-1}&=\exp\left[ \frac{r}{2}(\hat{x}\hat{p}+\hat{p}\hat{x})\right] 
\hat{x}
\exp\left[ -\frac{r}{2}(\hat{x}\hat{p}+\hat{p}\hat{x})\right] 
\nonumber \\ &
=\exp\left( -i r\right)  \hat{x} ,
\\
\hat{S}_r\hat{p}\hat{S}_r^{-1}&=\exp\left[ \frac{r}{2}(\hat{x}\hat{p}+\hat{p}\hat{x})\right] 
\hat{p}
\exp\left[ -\frac{r}{2}(\hat{x}\hat{p}+\hat{p}\hat{x})\right] 
\nonumber \\ &
=\exp\left(i r\right)  \hat{p} ,
\end{align}
\end{subequations}
which are derived from the commutation relation $[\hat{x},\hat{p}]=i$ and the Hadamard´s lemma \cite{rossman,hall,miller} $e^{g\hat{A}}\hat{B}e^{-g\hat{A}}=\hat{B}+g[\hat{A},\hat{B}]+\frac{g^2}{2!}[\hat{A},[\hat{A},\hat{B}]]+\dots$ when $g=\frac{r}{2}$; we obtain
\begin{equation}
\hat{\mathcal H}=\exp\left(2ir\right) \hat{p}^2
+\exp\left[ -ir\left( 2+\epsilon\right)+i\frac{\pi}{2}\epsilon \right]  \hat{x}^{2+\epsilon}.
\end{equation}
As $r$ is an arbitrary real constant, we choose it such that
\begin{equation}
\exp\left(2ir\right) =\exp\left[ -ir\left( 2+\epsilon\right)+i\frac{\pi}{2}\epsilon \right],
\end{equation}
and we get
\begin{equation}
r=\frac{\pi  (\epsilon +4 k)}{2 (\epsilon +4)},
\end{equation}
with $k$ an integer. Thus, the transformed Hamiltonian reads as
\begin{equation}
\hat{\mathcal{H}}_k= \exp\left(i\pi\frac{\epsilon+4k}{4+\epsilon} \right) 
\left( \hat{p}^2+\hat{x}^{2+\epsilon}\right),
\quad  k \in \mathbb{Z}.
\end{equation}
If we consider now a continuous  time Wick rotation, defining the imaginary time
\begin{equation}
\tau=t  \exp\left(i\pi\frac{\epsilon+4k}{4+\epsilon} \right)\, ,
\label{continuousWick}
\end{equation}
we get finally the Schrödinger equation
\begin{equation}
i\frac{\partial\psi(x,\tau)}{\partial \tau}=\left( \hat{p}^2+\hat{x}^{2+\epsilon}\right)\psi(x,\tau)
\end{equation}
with $\hat{p}^2+\hat{x}^{2+\epsilon}$ a Hermitian Hamiltonian.\\
The  Wick rotation  \eqref{continuousWick} belongs to a  family of general continuous rotations with the  form $t\rightarrow \exp(i\theta) t$, with $0\leq \theta\leq \pi/2$; therefore,  Wick rotations  may take us from Schrödinger-like equations to diffusion-like equations. They have also been proposed to develop a supersymmetric Euclidean theory from any supersymmetric Minkowski theory for Dirac spinors, thus producing an equivalence between bosons and fermions \cite{Waldron1,Waldron2}.

\section{Complex deformation of a general Hamiltonian} \label{sec5}
We may show that the very general $\mathcal{PT}$ symmetric, but non-Hermitian, Hamiltonian
\begin{equation}\label{2424}
\hat{H}= \hat{p}^2+V(i\hat{x})
\end{equation}
may be transformed with the non-unitary squeezing transformation 
$\hat{S}=\exp\left[\frac{\pi}{4}\left(\hat{x}\hat{p}+\hat{p}\hat{x}\right)\right]$ (which is the same used in the case of the complex deformed harmonic oscillator with $r=\pi/2$) to 
\begin{equation}
\hat{\mathcal{H}}= \exp\left(i \pi\right)\hat{p}^2+V(i\hat{x}e^{-i\frac{\pi}{2}}),
\end{equation}
which becomes the Hermitian Hamiltonian 
\begin{equation}
\hat{\mathcal{H}}=- \hat{p}^2+V(\hat{x}).
\end{equation}
As this Hamiltonian is Hermitian its evolution will be unitary and therefore  the usual techniques used in quantum mechanics may be applied. 

\section{Generalized Swanson oscillator} \label{sec6}
\begin{figure}[H]
\centering
	{\includegraphics[width=0.9\textwidth]{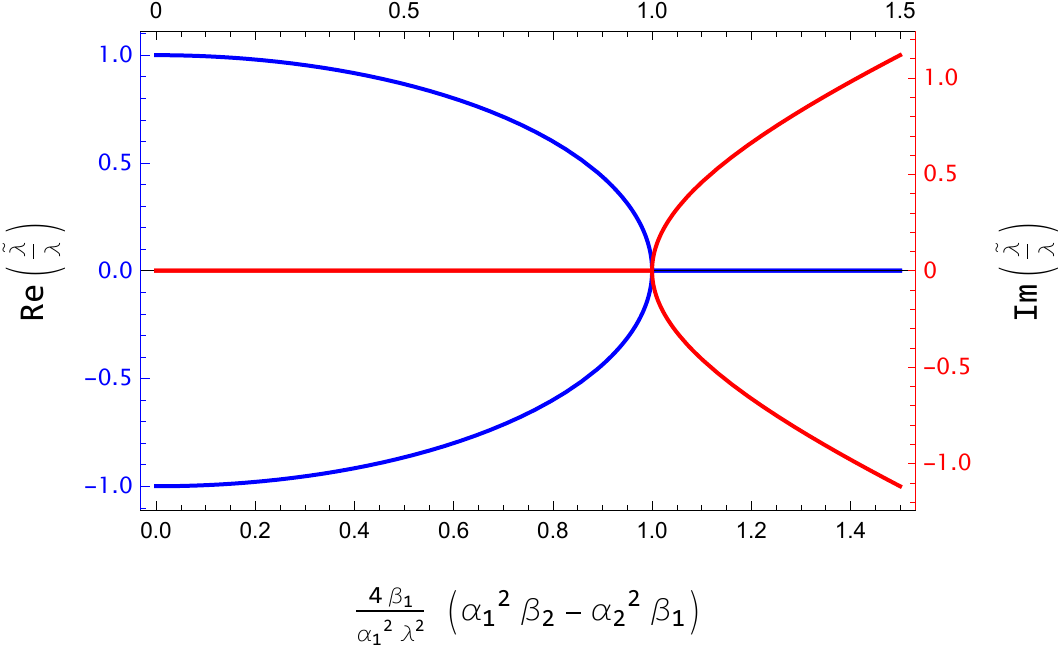}}
	\caption{Real and imaginary values of $\tilde{\lambda}/ \lambda$ as a function of $\frac{4 \beta_{1}}{\alpha_{1}^2 \lambda^2} \left( \alpha_{1}^2 \beta_{2}-\alpha_{2}^2 \beta_{1}\right)$. The function $\tilde{\lambda}/\lambda$ exhibits three different regimes depending on $\frac{4 \beta_{1}}{\alpha_{1}^2 \lambda^2} \left( \alpha_{1}^2 \beta_{2}-\alpha_{2}^2 \beta_{1}\right)$ is less, greater or equal to one.}
\label{f1}
\end{figure}

Let us now address another type of non-Hermitian system, a generalized version of the quadratic Swanson model $H_{SW}=\lambda \left( \hat{n} + 1/2\right)+\beta_{1} \hat{a}^{\dagger2}  + \beta_{2} \hat{a}^2$ \cite{15B,15.0B}; in effect, we are going to consider the Hamiltonian defined as
\begin{equation} \label{3.1}
\hat{H}_{GSW}=\lambda \left( \hat{n} + \frac{1}{2}\right)+ \alpha_{1} \hat{a}^{\dagger} + \alpha_{2} \hat{a} + \beta_{1} \hat{a}^{\dagger2}  + \beta_{2} \hat{a}^2, \quad \lambda,\alpha_{1}, \alpha_{2}, \beta_{1}, \beta_{2}  \in \mathbb R,
\end{equation}
where $\hat{a}^{\dagger} $ and $\hat{a}$ are the creation and annihilation operators of the standard harmonic oscillator and $\hat{n}=\hat{a}^\dagger \hat{a}$ is the number operator \cite{15.1B,15.2B,15.3B}; it is worthwhile to note that this Hamiltonian is a non-Hermitian version of a single-mode squeezed coherent harmonic oscillator. The presence of the linear terms makes the Hamiltonian in Eq.~\eqref{3.1} non-$\mathcal{PT}$-symmetric; this can be readily demonstrated applying the parity operator $\hat{P}:\hat{a}\rightarrow -\hat{a}$ and the time-reversal operator $\hat{T}: \hat{a}\rightarrow \hat{a}$, with similar transformations for $\hat{a}^{\dagger}$. For the particular case $\alpha_{1}=\alpha_{2}=0$, we get back to the original Swanson’s quadratic model, which has a $\mathcal{PT}$ symmetric non-Hermitian Hamiltonian \cite{15.4B,15.5B}. Conversely, when  $\alpha_{1}\neq 0$, $\alpha_{2}\neq 0$ but $\beta_{1}=\beta_{2}=0$, the system becomes a non-Hermitian forced harmonic oscillator. \\
In a sharp contrast with our recent examples which present a $\mathcal{PT}$ symmetry, here we are interested in translating the non-$\mathcal{PT}$-symmetric Hamiltonian Eq.~\eqref{3.2} to a $\mathcal{PT}$-symmetric one and show that it admits an equivalent Hermitian representation to the harmonic oscillator. Hence, we directly consider to solve the corresponding Schrödinger equation: $i\frac{d}{dt} \ket{\psi\left(t\right)}=\hat{H}\ket{\psi\left(t\right)}$; we will do that by means of three non-unitary transformations. We begin by introducing the transformation $\hat{\eta}_{1}=\exp\left[\frac{\hat{n}}{2} \ln\left(\frac{\alpha_{2}}{\alpha_{1}}\right) \right]$ with $\alpha_{1} \neq 0$ and $\alpha_{2} \neq 0$; the transformed Hamiltonian becomes
\begin{align} \label{3.2}
\hat{H}_1=\hat{\eta}_{1}\hat{H}_{GSW}\hat{\eta}_{1}^{-1}=
\lambda \left(\hat{n} + \frac{1}{2}\right) + \sqrt{\alpha_{1} \alpha_{2}} \left(\hat{a}^{\dagger} + \hat{a}\right) 
+\left( \frac{\alpha_{2}\beta_{1}}{\alpha_{1}} \hat{a}^{\dagger2}+ \frac{\alpha_{1}\beta_{2}}{\alpha_{2}} \hat{a}^{2}\right).
\end{align}
Note that this transformed Hamiltonian is non-Hermitian and non-$\mathcal{PT}$-symmetric. Remark also that the application of the operator $\hat{\eta}_{1}$ makes that the linear terms in $\hat{a}$ and $\hat{a}^{\dagger}$ can be expressed as a sum of both operators multiplied by the factor $\sqrt{\alpha_{1}\alpha_{2}}$.\\
We get rid of the linear terms by means of the simple non-unitary transformation $\hat{\eta}_{2}=\exp\left(\gamma_{1} \hat{a}^{\dagger}-\gamma_{2}\hat{a} \right)$, where $\gamma_{1}$ and $\gamma_{2}$ are two unknown parameters to be determined. Remark that $\hat{\eta}_2$  translates into the usual displacement operator once $\gamma_{1}=\gamma_{2}$; then, the application of $\hat{\eta}_2$ to the Hamiltonian $\hat{H}_1$, Eq.~\eqref{3.2}, yields to
\begin{align} \label{3.3}
\hat{H}_2=\hat{\eta}_{2}\hat{H}_1\hat{\eta}_{2}^{-1}=&
\lambda \left( \hat{n} + \frac{1}{2} \right) + \frac{\alpha_{2}\beta_{1}}{\alpha_{1}} \hat{a}^{\dagger2}+ \frac{\alpha_{1}\beta_{2}}{\alpha_{2}} \hat{a}^{2} 
+\left(\sqrt{\alpha_{1}\alpha_{2}}-\gamma_{1} \lambda  - \frac{2\gamma_{2}\alpha_{2} \beta_{1}}{\alpha_{1}}\right) \hat{a}^{\dagger} 
\nonumber\\ &
+\left(\sqrt{\alpha_{1}\alpha_{2}}-\gamma_{2} \lambda  - \frac{2\gamma_{1}\alpha_{1} \beta_{2}}{\alpha_{2}}\right) \hat{a} 
+\lambda \gamma_{1} \gamma_{2} -\sqrt{\alpha_{1} \alpha_{2}}\left(\gamma_{1} + \gamma_{2}\right)
+ \frac{\gamma_{1}^2 \alpha_{1} \beta_{2}}{ \alpha_{2}} + \frac{\gamma_{2}^2 \alpha_{2} \beta_{1}}{ \alpha_{1}}.
\end{align}
In general, this transformed Hamiltonian is non-Hermitian and non-$\mathcal{PT}$-symmetric. We choose the parameters $\gamma_{1}$ and $\gamma_{2}$ of the $\hat{\eta}_{2}$ transformation in such a way that the coefficients of the linear terms are zero; i.e., through the equations
\begin{align} \label{3.4}
\gamma_{1} \lambda -\sqrt{\alpha_{1}\alpha_{2}} + \frac{2\gamma_{2}\alpha_{2} \beta_{1}}{\alpha_{1}}=0, 
\qquad
\gamma_{2} \lambda -\sqrt{\alpha_{1}\alpha_{2}} + \frac{2\gamma_{1}\alpha_{1} \beta_{2}}{\alpha_{2}}=0,
\end{align}
which gives
\begin{align} \label{3.5}
\gamma_{1}=\frac{\alpha _2}{\sqrt{\alpha _1 \alpha _2}}
\frac{\alpha _1 \lambda -2 \alpha _2 \beta _1}{\lambda ^2-4 \beta _1 \beta _2},
\qquad
\gamma_{2}=\frac{\alpha _1}{\sqrt{\alpha _1 \alpha _2}}
\frac{\alpha _2 \lambda -2 \alpha _1 \beta _2}{\lambda ^2-4 \beta _1 \beta _2}.
\end{align}
Substituting these values in the Hamiltonian $\hat{H}_2$, Eq.~\eqref{3.3}, it takes the reduce form
\begin{equation} \label{3.6}
\hat{H}_2=\lambda \left(\hat{n}+\frac{1}{2}\right)+ \frac{\alpha_{2}\beta_{1}}{\alpha_{1}} \hat{a}^{\dagger2}+ \frac{\alpha_{1}\beta_{2}}{\alpha_{2}} \hat{a}^{2} + \delta,
\end{equation}
where
\begin{equation} \label{3.7}
\delta= \frac{\alpha^2_{1}\beta_{2}+ \alpha^2_{2}\beta_{1}-\lambda \alpha_{1}\alpha_{2} }{\lambda^2-4 \beta_{1} \beta_{2}}.
\end{equation}
This Hamiltonian is non-Hermitian but it is $\mathcal{PT}$-symmetric, and possess the same structure as the Hamiltonian of the Swanson oscillator.\\
In order to solve the Schrodinger equation associated with the Hamiltonian $\hat{H}_2$, we make the third and final transformation $\hat{\eta}_{3}=\exp\left( \zeta \hat{a}^2 \right)$. If we select
\begin{equation}
\zeta=\frac{-\lambda \alpha _1 + \sqrt{\alpha _1^2 \lambda ^2-4 \alpha _1^2 \beta _1 \beta _2+4 \alpha _2^2 \beta _1^2}}{4 \alpha _2 \beta _1},
\end{equation}
the transformed Hamiltonian is
\begin{align}\label{ham3}
\hat{H}_3=\hat{\eta}_{3}\hat{H}_2\hat{\eta}_{3}^{-1}=
\tilde{\lambda} \left( \hat{n}+\frac{1}{2} \right) 
+\frac{\alpha _2 \beta _1}{\alpha _1}\left(\hat{a}^{\dagger 2}+\hat{a}^2\right)+\delta,
\end{align}
where
\begin{equation}
\tilde{\lambda }= \lambda \sqrt{1-\frac{4 \beta_{1}}{\alpha_{1}^2 \lambda^2} \left( \alpha_{1}^2 \beta_{2}-\alpha_{2}^2 \beta_{1}\right)}.
\end{equation}
The Hamiltonian $\hat{H}_3$, Eq.~\eqref{ham3}, is Hermitian if and only if $\frac{4 \beta_{1}}{\alpha_{1}^2 \lambda^2} \left( \alpha_{1}^2 \beta_{2}-\alpha_{2}^2 \beta_{1}\right) < 1$; in this case, the transformation $\hat{\eta}_{3}=\exp\left( \zeta \hat{a}^2 \right)$ is non-unitary as $\zeta$ can take two real values from the different signs of its  square-root expression. This can be appreciated more clearly in Fig.~\ref{f1}, where in the regime $\frac{4 \beta_{1}}{\alpha_{1}^2 \lambda^2} \left( \alpha_{1}^2 \beta_{2}-\alpha_{2}^2 \beta_{1}\right)<1$, the model has real values (solid blue line) corresponding to the unbroken ${\mathcal PT}$ symmetry. On the other hand, the situation is rather different if $\frac{4 \beta_{1}}{\alpha_{1}^2 \lambda^2} \left( \alpha_{1}^2 \beta_{2}-\alpha_{2}^2 \beta_{1}\right) > 1$; the Hamiltonian $\hat{H}_3$, Eq.~\eqref{ham3}, is not hermitian anymore, since $\tilde{\lambda}$ appears in conjugate pairs of purely imaginary values, and the values of $\zeta$ become complex conjugate of each other; above abrupt transition indicates a broken ${\mathcal {PT}}$-symmetry of the system. In the case $\frac{4 \beta_{1}}{\alpha_{1}^2 \lambda^2} \left( \alpha_{1}^2 \beta_{2}-\alpha_{2}^2 \beta_{1}\right)=1$, the eigenvalues coalesce and we are at the exceptional point; there we have only one eigenvalue. Thus, we restrict ourselves to the range of parameters where $\zeta$ is real and positive (positive sign of the square root), where the Hamiltonian Eq.~\eqref{ham3} is Hermitian and we can guarantee real eigenvalues. This is also consistent with the conjecture of Bender\cite{bender98,bender02,12} and in exact accordance with the typical Swanson’s Hamiltonian which admits a purely real positive spectrum.\\
The exact solution of the Schrödinger equation corresponding to the Hamiltonian \eqref{ham3} is
\begin{equation} \label{13}
 \ket{\psi_3\left(t\right)}=\exp\left(-i \delta t \right) \exp \left(-2 i \frac{\alpha_2 \beta_1}{\alpha_1} \hat{\mathscr{H}} t \right)  \ket{\psi_3\left(0\right)},
\end{equation}
where
\begin{equation}
\hat{\mathscr{H}}=\hat{K}^{+} + \frac{\alpha_{1}\tilde{\lambda}}{\alpha_{2} \beta_{1}} \hat{K}^{0}+ \hat{K}^{-},
\end{equation}
being $\hat{K}^{+}=\frac{1}{2}\hat{a}^{\dagger2}$, $ \hat{K}^{0}=\frac{1}{2}\left(\hat{n} +\frac{1}{2}\right)$ and $ \hat{K}^{-}=\frac{1}{2}\hat{a}^{2}$. The commutation relations of these operators are $\left[\hat{K}^{+}, \hat{K}^{-} \right]=-2\hat{K}^0$, $\left[\hat{K}^{0}, \hat{K}^{\pm} \right]=\pm \hat{K}^{\pm}$, and they are the generators of the su(1,1) algebra \cite{15.6B,15.7B,15.8B,15.9B,15.10B,15.11B,15.12B}. Nonetheless, one may easily convert the Hamiltonian $\hat{H}_3$ into the diagonal Hamiltonian, $\hat{\tilde{H}}_{3}=\hat{S}^{-1} \hat{H}_{3} \hat{S}=2\sqrt{\lambda^2-4\beta_{1}\beta_{2}}\hat{K}^{0} + \delta$ by applying the squeeze-like transformation
\begin{equation}
\hat{S}=\exp\left[-\frac{1}{2}\arctanh\left(\frac{2\alpha_{2}\beta_{1}}{\alpha_{1}\tilde{\lambda}}\right)\left(\hat{K}^{+}-\hat{K}^{-}\right) \right].
\end{equation}
Since the eigenstates of $\hat{K}^{0}$ are $\ket{n}$ with eigenvalues $\frac{1}{2}\left(n+\frac{1}{2}\right)$, then, the Hamiltonian $\tilde{H}_{3}$ corresponds to the harmonic oscillator with frequency $\sqrt{\lambda^2-4\beta_{1}\beta_{2}}$. 
Consequently, we at once give to the energy spectrum of our original Hamiltonian \eqref{3.1} from the eigenequation of the Hamiltonian $\hat{\tilde{H}}_{3}$, i.e, $\hat{\tilde{H}}_{3}\ket{n}=E_{n}\ket{n}$, where
\begin{equation}
E_{n}=\sqrt{\lambda^2-4\beta_{1}\beta_{2}}\left(n+\frac{1}{2}\right) + \frac{\alpha^2_{1}\beta_{2}+ \alpha^2_{2}\beta_{1}-\lambda \alpha_{1}\alpha_{2} }{\lambda^2-4 \beta_{1} \beta_{2}},  
\end{equation}
being real as long as $\lambda^2-4\beta_{1} \beta_{2} \gg 0$. Meanwhile the eigenstates of $\hat{H}_{GSW}$ are given by $\ket{\tilde{n}}=\hat{\eta}_{1}^{-1}\hat{\eta}_{2}^{-1}\hat{\eta}_{3}^{-1}\hat{S}\ket{n}$, where $\hat{\eta}_{1}^{-1}, \hat{\eta}_{2}^{-1}, \hat{\eta}_{3}^{-1}$and $\hat{S}$ act like intertwining operators, in the sense that they transform an eigenstate of $\hat{\tilde{H}}_{3}$ to eigenstate of $\hat{H}_{GSW}$ with the same spectrum through these operators. In other words, the eigenfunctions of $\hat{H}_{GSW}$ can be derived from the eigenfunctions of the harmonic oscillator.\\
Let us finish this section by mentioning that $\hat{H}_{GSW}$ can be recast in terms of position and momentum operators by the well-known relationships, $\hat{a}=\sqrt{\frac{m \omega}{2 }}\left(\hat{x}+i\frac{\hat{p}}{m \omega}\right)$ and $\hat{a}^{\dagger}=\sqrt{\frac{m_{0} \omega_{0}}{2 }}\left(\hat{x}-i\frac{\hat{p}}{m_{0} \omega_{0}}\right)$, being $\omega$ and $m$ the constant frequency and the mass; this leads us to the Hamiltonian form
\begin{equation}
    \hat{H}_{GSW}=\nu_{1}\hat{p}^2+ \nu_{2}\hat{x}^2+i\nu_{3}\left(\hat{x}\hat{p}+\hat{p}\hat{x}\right)+i\nu_{4} \hat{p}+\nu_{5}\hat{x},
\end{equation}
with energy spectrum
\begin{equation}
 E_{n}=\sqrt{\nu_{1}\nu_{2}+\nu^2_{3}} \left(2n+1\right) + \frac{\nu^2_{4}\nu_{2}-\nu^2_{5}\nu_{1}-2\nu_{3}\nu_{4}\nu_{5}}{4\left(\nu_{1}\nu_{2}+\nu^2_{3}\right)},
\end{equation}
being 
\begin{align}
    \nu_{1}=&\frac{1}{2 m\omega}\left(\lambda-\beta_{1}-\beta_{2}\right),
    \quad
    \nu_{2}=\frac{m \omega}{2} \left(\lambda + \beta_{1}+\beta_{2}\right),
    \quad
    \nu_{3}=\frac{1}{2}\left(\beta_{2}-\beta_{1}\right),
    \nonumber  \\ 
    \nu_{4}=&\frac{1}{\sqrt{2 m \omega}}\left(\alpha_{2}-\alpha_{1}\right)
    \qquad \;
    \nu_{5}=\sqrt{\frac{m \omega}{2}}\left(\alpha_{1}+\alpha_{2}\right).
\end{align}
Using the above Ansatz , one can get the corresponding solution for the Schrödinger equation governed by the Hamiltonian $\hat{H}_{GSW}$ in the time dependent case as well as in the time-independent scenario, as already reported in \cite{15.121B}.

\subsection{Simple photonic lattice analog} \label{subsec6}
\begin{figure}[H]
	\centering
	{\includegraphics[width=0.8\textwidth]{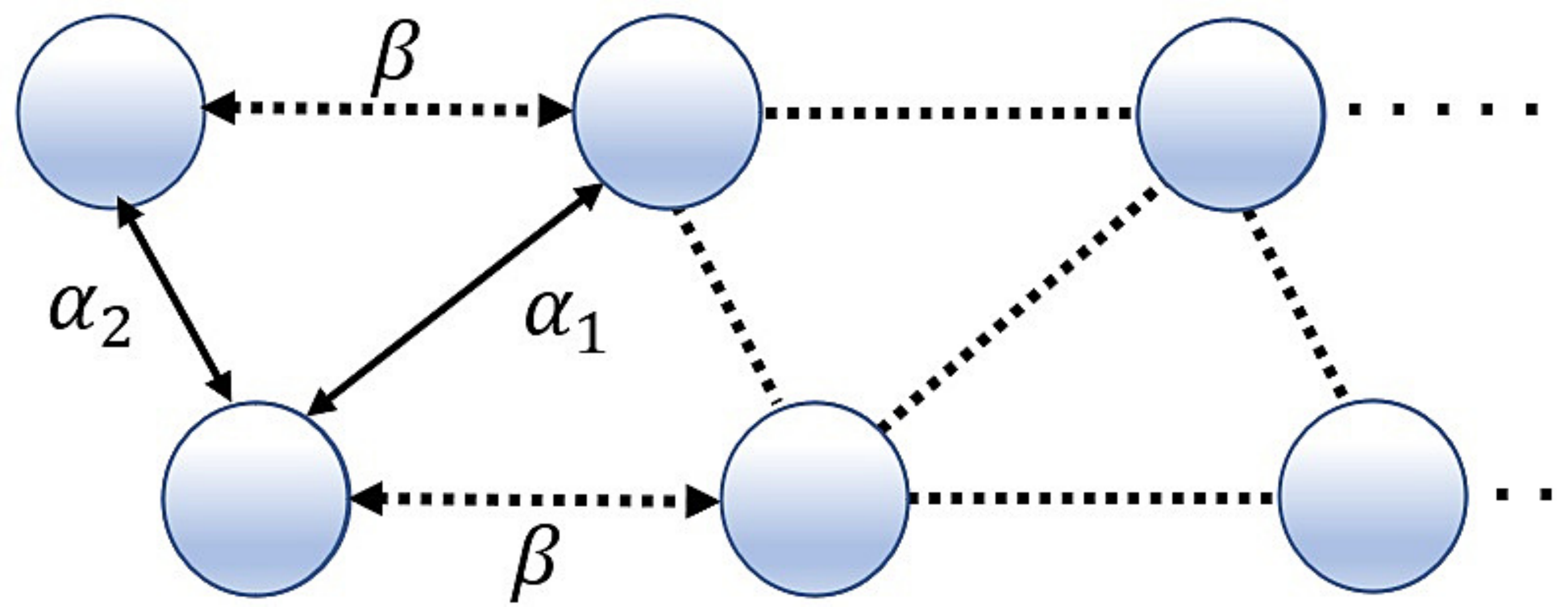}}
	\caption{Sketch of zigzag waveguide array with nearest, $\left(\alpha_{1}, \alpha_{2}\right)$, and next-nearest,$\left(\beta\right)$, neighbor hopping.}
	\label{f2}
\end{figure}

Lastly, a simple analog of the previous model for the particular situation $\beta_{1}=\beta_{2}$ can be realized by using a zigzag waveguide with nonuniform nearest-neighbor hopping that depend on the square root of the site number. The light evolution in the non-Hermitian lattice satisfies the differential equation set
\begin{align}  \label{3.13}
i \frac{d\Psi_{n}(Z) }{d Z}& +  \lambda \left(n + \frac{1}{2} \right) \Psi_{n}(Z) + \alpha_{1} \sqrt{n} \;  \Psi_{n-1}(Z) 
+ \alpha_{2} \sqrt{n+1}  \;  \Psi_{n+1}(Z)
\nonumber\\ & 
+\beta \sqrt{n(n-1)}  \;  \Psi_{n-2}(Z) 
+  \beta \sqrt{(n+1)(n+2)}  \;  \Psi_{n+2}(Z) = 0, \quad n=0,1,2,...,
\end{align}
where $\Psi_{n}(Z)$ is the complex field amplitude in the site $n$ at the dimensionless propagation distance $Z$, and we adopt the convention that $\Psi_{n}(Z)=0 $ for $n<0$; the term $\lambda n$ is the refractive index and it varies gradually with the site number. The parameters $\alpha_{1}$ and $\alpha_{2}$ denote the left and right nearest-neighbor hopping and $\beta$ represents the next-nearest interaction. In order to describe the evolution of nonclassical light in the above waveguide system, it is convenient to use a simplified notation in which each single-mode waveguides are arranged in a vector given by $\ket{\psi\left(Z\right)}=\sum_{n=0}^{\infty} \Psi_{n}(Z) \ket{n}$, where $\ket{n}$ plays an analogous role to Fock states. In this form, Eq.\eqref{3.13} can be rewritten into a rather simple and suggestive  Schrödinger-like equation form,
\begin{align}  \label{3.14}
   i\frac{d\ket{\psi\left(Z\right)}}{dZ} =&-\left[ \lambda \left( \hat{n} + \frac{1}{2}\right) + \alpha_{1} \hat{a}^{\dagger} + \alpha_{2} \hat{a} 
   + \beta \left( \hat{a}^{\dagger2}  + \hat{a}^2 \right)\right] \ket{\psi\left(Z\right)},
\end{align}
whose Hamiltonian is the same as \eqref{3.1}, when $\beta_{1}=\beta_{2}$. In fact, one can easily check that substituting this vector proposal into Eq.~\eqref{3.14} yields the waveguide system given by Eq.~\eqref{3.13}.\\ 
A flagship example of how the lattice can be engineered is given in \cite{15.13B}; there, the zigzag shape consists of two interleaved waveguides (see Fig.~\ref{f2}) where one layer forms a scalene triangle with two adjacent waveguides in the down layer; such scalene configuration gives rise to unequal cross coupling nearest-neighbor, $\alpha_{1}$ and $\alpha_{2}$, whereas the next neighbor hopping, $\beta$, is due to the coupling of waveguides in the same layer.
Further, the choice of a different geometrical setting, such as a one-dimensional linear chain of waveguides arrays, leads to the next-nearest interaction effects starting to become completely insignificant, $\beta=0$; as a result, the problem is reduced to the Non-Hermitian Glauber–Fock lattice \cite{16.A,16C,16B} with a transverse ramp of refractive index \cite{17B} when $\lambda \neq 0$. Finally, for the case $\alpha_{1}=\alpha_{2}=\alpha$ and $\beta_{1}=\beta_{2}=\beta$, the Eq~\eqref{3.13} turns out to be the zigzag lattice reported in \cite{15.8B}.

\section{Conclusion}\label{sec7}
We have tried to show, that by using non-unitary transformations, non-Hermitian Hamiltonians may be produced. We have used in most of our results a non-unitary "squeezed operator". However, such operator may be greatly extended; for instance, consider the time dependent harmonic oscillator Hamiltonian (we set the frequency equal to one for simplicity)
\begin{equation}
    H(t)= \hat{a}^{\dagger}\hat{a}+\frac{\lambda}{\sqrt{2}} (\hat{a}e^{i t}+\hat{a}^{\dagger}e^{-i t}),
\end{equation}
with the driving amplitude given by $\frac{\lambda}{\sqrt{2}} $. The simple unitary transformation, $e^{-i\hat{a}^{\dagger}\hat{a}}$ may be used to take such Hamiltonian  to the time independent one
\begin{equation}
    H=\frac{\lambda}{\sqrt{2}}\hat{x}. 
\end{equation}
This Hamiltonian has  position as eigenstates eigenfunctions that, unfortunately, are not normalized and therefore they are not proper wavefunctions. We may produce the nonunitary transformation \cite{Mielnik} $T=e^{-{\frac{\hat{p}^2}{2}}}$ such that we obtain the non-Hermitian Hamiltonian
\begin{equation}
    H_T=THT^{\dagger}=\frac{\lambda}{\sqrt{2}}(\hat{x}+i\hat{p})=\lambda\hat{a},
\end{equation}
whose eigenfunctions are coherent states, {\it i.e.}, properly normalized wavefunctions. Both Hamiltonians, $H$ and $H_T$ {\it share the same eigenvalues because they are related by a transformation}, but the first has unnormalized eigenfunctions while the last has normalized ones. The transformation, $T$, has as argument of the exponential the operator $\hat{p}^2$ that form an algebraic group, namely, $SU(2)$, with $\hat{x}^2$ and $\hat{x}\hat{p}+\hat{p}\hat{x}$. This last operator we have used in Section 4 to produce a Hermitian Hamiltonian from the paradigmatic example of $\mathcal {PT}$ symmetric infinite dimensional Hamiltonians. The fact that the operator involved in such transformation belongs to the $SU(2)$ group, just as the one used in this Section, to transform the position operator into the annihilation operator makes it clear that there are no concerns over the domain in which solutions are valid, but only the usual concerns related to the use of non-Hermitian Hamiltonians that obviously produce not proper wavefunctions as they do not conserve probability. \\
Many more nonunitary transformations may be used to produce non-Hermitian Hamiltonians from Hermitian ones, for instance, $R_x=\exp(g[\hat{x}])$ or $R_p=\exp(f[\hat{p}])$ where $g$ and $f$ are in general a complex functions.\\
Moreover, we have applied such non-unitary transformations even to the case where an effective potential may be considered a function of position and momentum. In particular, note that the Swanson Hamiltonian (\ref{3.1}) is more general than the Hamiltonian (\ref{2424}), in the sense that it has terms of the form $\hat{a}^2$, and therefore it implies a potential that not only depends on the position operator, $\hat{x}$, but also on the momentum operator, $\hat{p}$. We have associated the complex time produced by such non-unitary transformations to Wick rotations.\\

\section{Acknowledgments}
B.M. Villegas-Martínez wishes to express his gratitude to CONACyT as well as to the National Institute of Astrophysics, Optics and Electronics (INAOE) for financial support.

\section{Author Contributions} 
All authors have contributed equally to this work. All authors have read and agreed to the published version of the manuscript.

\section{Conflicts of Interest}
The authors declare no conflict of interest.

\section{Data Availability Statement}
No Data associated in the manuscript

\end{document}